\documentclass{article}
\usepackage[utf8]{inputenc}
\usepackage{subcaption}
\usepackage{graphicx}
\usepackage{multirow}
\usepackage{wrapfig}
\usepackage[numbers]{natbib}  
\usepackage{booktabs}
\usepackage[left]{lineno}
\usepackage[hyphens]{url}
\usepackage[a4paper, left=3cm, right=3cm, top=2cm, bottom=2cm]{geometry}

\usepackage{hyperref}
\usepackage[hyphenbreaks]{breakurl}
\usepackage{breakcites}
\hypersetup{
    colorlinks=true,
    linkcolor=blue,
    filecolor=magenta,      
    urlcolor=cyan,
    }

\title{Global birdsong embeddings enable superior transfer learning for bioacoustic classification}

\author{Burooj Ghani$^{1\star}$, Tom Denton$^{2\star}$, Stefan Kahl$^{3, 4}$, Holger Klinck$^{3}$\\
    \small $^{1}$Naturalis Biodiversity Center, $^{2}$Google Research\\
    \small $^{3}$K. Lisa Yang Center for Conservation Bioacoustics, Cornell Lab of Ornithology, Cornell University\\
    \small $^{4}$Chemnitz University of Technology
}

\begin{document}

\maketitle

\renewcommand{\thefootnote}{\fnsymbol{footnote}}
\footnotetext[1]{The two authors contributed  equally to this paper and share first authorship. \newline  (Email: burooj.ghani@naturalis.nl; tomdenton@google.com)}
\begin{abstract} 
\noindent 
Automated bioacoustic analysis aids understanding and protection of both marine and terrestrial animals and their habitats across extensive spatiotemporal scales, and typically involves analyzing vast collections of acoustic data. With the advent of deep learning models, classification of important signals from these datasets has markedly improved. These models power critical data analyses for research and decision-making in biodiversity monitoring, animal behaviour studies, and natural resource management. However, deep learning models are often data-hungry and require a significant amount of labeled training data to perform well. While sufficient training data is available for certain taxonomic groups (e.g., common bird species), many classes (such as  rare and endangered species, many non-bird taxa, and call-type) lack enough data to train a robust model from scratch.
This study investigates the utility of feature embeddings extracted from audio classification models to identify bioacoustic classes other than the ones these models were originally trained on. We evaluate models on diverse datasets, including different bird calls and dialect types, bat calls, marine mammals calls, and amphibians calls.  The embeddings extracted from the models trained on bird vocalization data consistently allowed higher quality classification than the embeddings trained on general audio datasets. The results of this study indicate that high-quality feature embeddings from large-scale acoustic bird classifiers can be harnessed for few-shot transfer learning, enabling the learning of new classes from a limited quantity of training data. Our findings reveal the potential for efficient analyses of novel bioacoustic tasks, even in scenarios where available training data is limited to a few samples.

\end{abstract}

\section{Introduction}

Bioacoustic analysis provides a rich window into biodiversity, animal behavior and ecosystem health. Passive acoustic monitoring (PAM) in particular has become a widely used tool for wildlife conservation. PAM uses battery-operated autonomous recording devices (ARUs) that collect vast amounts of acoustic data, containing a wealth of information about biological, geophysical, and anthropogenic activities in the deployment area. It allows researchers to study and protect animals and their habitats non-invasively at ecologically-relevant temporal and spatial scales~\citep{sugai2019terrestrial}. PAM involves recording sound in nature and has been used to study a wide range of species, including whales and dolphins~\citep{estabrook2022dynamic, fouda2018dolphins}, pinnipeds~\citep{van2010acoustic, crance2022year}, birds~\citep{wood2019acoustic, symes2022analytical}, insects~\citep{symes2022estimation, mankin2011perspective}, fish~\citep{rountree2006listening, desidera2019acoustic}, frogs~\citep{nelson2017seasonal, measey2017counting}, and terrestrial mammals~\citep{clink2023workflow, swider2022passive}. In recent years, many automated deep learning-based analysis tools have been developed that are now commonly used to analyze long-term acoustic data efficiently~\citep{stowell2022computational}. By utilizing these tools, researchers can automatically detect and categorize animal vocalizations, saving them a significant amount of time and effort and facilitating the investigation of less researched species~\citep{brunk2023quail}. However, the development of these tools typically depends on the availability of well-annotated training data. Obtaining sufficient training data can be a major challenge. While there are sufficient amounts of training data available for some taxonomic groups, including common bird species (e.g., through community collections like Xeno-canto\footnote[2]{https://xeno-canto.org} or the Macaulay Library\footnote[3]{https://www.macaulaylibrary.org}), training data is often lacking for rare and endangered species, which are often the prime target of conservation efforts~\citep{stowell2019automatic}. In addition, traditional approaches to species-level classification may not be suitable for all applications. For example, a fixed set of classes may not be desirable in cases where researchers are interested in the fine-grained classification of vocalizations, such as identifying specific call types rather than simply identifying the presence or absence of a species~\citep{GhaniMachineLA}. Call types and the associated behaviors (e.g., foraging or breeding) can provide critically important cues on habitat use and inform, for example, land management decisions.

One way to address the challenge of data deficiencies is to utilize learned feature embeddings for \emph{few-shot transfer learning}. In the context of machine learning, \emph{feature embeddings} are vectors obtained from some intermediate layer of a trained machine learning model~\citep{stowell2022computational}. Since 2014, numerous studies have found that embeddings from pre-trained models can allow more efficient learning of novel tasks~\citep{oquab2014learning, yosinski2014transferable, chollet2017limitations}.
\emph{Few-shot learning} refers to a wide array of methods attempting to produce strong models with little training data, inspired by the apparent human ability to learn new classes from a handful of examples~\citep{wang2020generalizing}. Feature embeddings often figure prominently in few-shot learning designs, allowing \emph{transfer} of learned features to new tasks and domains.

High-quality feature embeddings offer several benefits over traditional approaches to species-level classification. First, feature embeddings can help to differentiate between classes of acoustic events that are very similar and differ only in subtle details. For instance, songbirds can display local variations (also called dialects) in their song patterns, which may lead to slight differences in note sequences~\citep{catchpole2003bird}. Feature embeddings can capture these nuances and enable more precise classification. Additionally, embeddings facilitate transfer learning between species, enabling researchers to train models on data from more commonly occurring or extensively studied species and then apply that knowledge to a target species, which may have insufficient training data. This approach also saves researchers time and effort that would otherwise be needed to train a dedicated classifier from scratch while enhancing the accuracy of classification results. Furthermore, cross-taxa classification based on feature embeddings is also possible when such embeddings can generalize across acoustic domains and events.

Because the relevant features for different problems may vary, we hypothesize that models trained on a problem closely related to the target problem will often outperform models trained on very different problems. In fact, the recent HEAR Benchmark competition found that no single model dominated across event detection, music transcription, and speech recognition tasks~\citep{turian2022hear}.
However, as mentioned earlier, many problems lack sufficient data for training a robust classifier from scratch. In these cases, re-using the feature embeddings from a pre-trained model allows learning the new task efficiently, so long as the embeddings are sufficiently relevant.

In this study, we investigate the use of various large-scale acoustic classifiers to produce feature embeddings that can be used to perform fine-grained classification of bird calls and dialect types, and out-of-scope but related identification of acoustic events (non-bird animal calls) that these models have not been trained on. Furthermore, we include in our analysis classifiers that are either trained on AudioSet dataset~\citep{gemmeke2017audio} (a broad spectrum of audio data extracted from YouTube clips) or on extensive datasets of bird vocalisations from around the world. In doing so, we are able to compare the effectiveness of these embeddings derived from different classifiers, evaluating their capacity to generalize and detect a variety of bioacoustic events.

The paper aims to provide a simple method for species-agnostic classification across taxonomic groups by leveraging transfer learning capabilities of selected classifiers. The effectiveness of the approach is demonstrated by evaluating on a diverse set of data sources covering birds, bats, marine mammals, and amphibians. Overall, our study suggests that the proposed approach can help to advance automated analysis in passive acoustic monitoring by solving the problem of species and call type recognition in low- and medium-data regimes. The use of transfer learning capabilities of selected classifiers provides a practical and effective way to classify a wide range of acoustic events across different taxa and can help to improve the accuracy and efficiency of PAM analysis efforts. 

Our approach -- utilizing fixed, pre-trained embeddings for novel problems -- also suggests a more efficient workflow for large-scale bioacoustic data sets. Large PAM deployments may accumulate tens to hundreds of terabytes of data during a single field season~\citep{oswald2022collection}. This makes model inference tasks especially time-consuming and potentially expensive. Given a model which produces generally useful feature embeddings, the practitioner may embed their entire data set once and then use the pre-computed embeddings for a wide range of subsequent analysis tasks. Training and inference with small models over fixed embeddings are much faster than training entirely new models: Training a high-quality classifier from scratch can take many days of GPU time, but training small linear classifiers over fixed embeddings, which we discuss in this paper, can take less than a minute to train on a modern workstation. This allows fast experimentation with different analysis techniques and quickly iterating with human-in-the-loop active learning techniques.

\subsection{Relationship to Previous Work}

Previous works have investigated transfer learning for novel bioacoustic tasks, including the use of pre-trained global bird embeddings (\citep{mcginn2023feature}, \citep{tolkova2021parsing, boudiaf2023search}). Many of these prior works focus on a single use-case, leaving the question of breadth of generalization unanswered.

Other studies on bioacoustic transfer learning use only embeddings from VGGish~\citep{vggish}, a general audio model. This is an older model, and even recent bioacoustic studies (\citep{sethi2022soundscapes, hagiwara2022beans, heath2021index, ccoban2020transfer}) continue to use it, despite long being surpassed on general audio benchmarks. In this work, we provide comparisons of bioacoustic transfer learning from VGGish to more recent SOTA models such as AudioMAE (a self-supervised transformer), YAMNet, and PSLA (more recent convolutional models).

Our study provides a far more complete comparison of global bird embeddings against general audio models, across a diverse array of datasets, demonstrating robust generalization.

\section{Methods}

\begin{figure}[t!]
  \centering
  \includegraphics[width=1.1\textwidth,trim={1.2cm 4.5cm 0 4.5cm 0},clip]{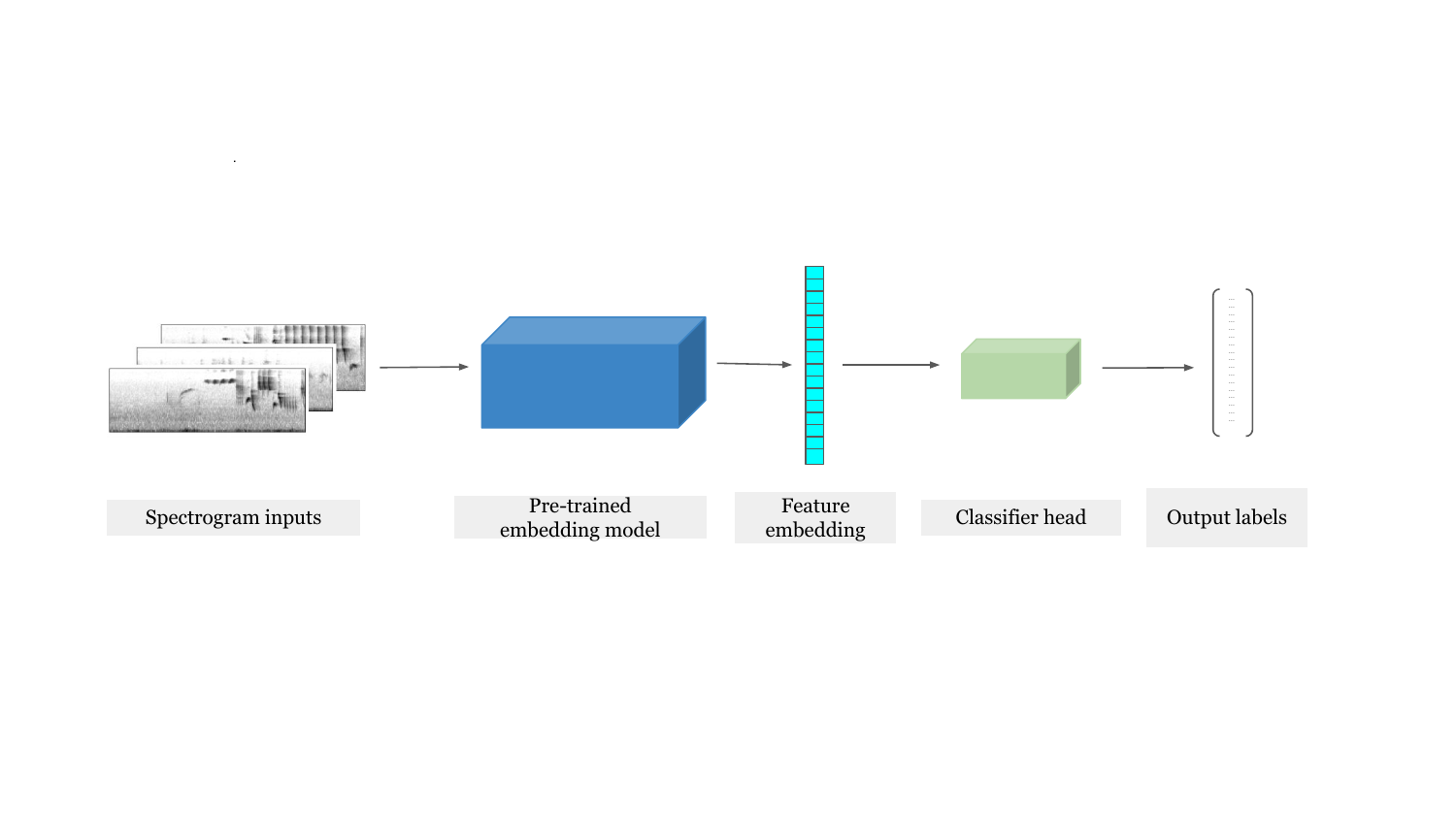}
  \caption{Workflow illustrating the process of downstream classification for various bioacoustic tasks. Spectrograms are processed by pre-trained models, producing embeddings. These embeddings are then passed to a classifier head—a fully-connected feed forward neural network—which is trained for final classification.}
  \label{fig:pipeline}
\end{figure}

In this work, we focus on the extraction of feature embeddings from six CNN models and one transformer model, described below in Section~\ref{sec:models}. These \emph{embedding models} are trained on either general YouTube data or global data sets of bird vocalizations. All models map spectrograms (visual representation of sound) to their class labels. Using logistic regression, we train classifiers on the feature embeddings extracted from each model as described in Section~\ref{sec:linprobe_methods}. Fig.~\ref{fig:pipeline} provides an overview of the classification pipeline we employed for our experiments. 

Spectrograms serve as the input data for our framework. The pre-trained embedding model, which is essentially the large-scale classifier without the classifier head, processes the spectrograms and produces an embedding. The embedding can be seen as a compact representation capturing the salient features of the input. This embedding is then forwarded to the classifier head, which is implemented as a fully connected layer. The classifier head applies a linear transformation to the embedding, followed by a sigmoid function to obtain class probabilities, and is trained via standard logistic regression. In summary, this architecture, comprising the embedding model, fully connected layer, and sigmoid activation, enables the extraction of relevant features from spectrograms and the subsequent generation of probability estimates for downstream classification purposes. 

By employing simple logistic regression, we are able to judge the direct utility of each model's pre-trained embedding to a range of problems. Additionally, we save an immense amount of training effort by pre-computing the embeddings for each dataset.

\subsection{Linear Probes on Feature Embeddings}
\label{sec:linprobe_methods}

In this work, we consider the case of keeping the entire pre-trained embedding frozen and learning a single linear layer for the new tasks. This method is essentially a \emph{linear probe} of the selected embeddings, which allows assessment of the availability of desired task-specific information in the embeddings~\citep{alain2016understanding}. If the downstream task can be performed well with just a linear layer on top of these embeddings, it suggests that the embeddings already have the necessary information encoded in them. 

For each pairing of model and data set, we first calculate the model embeddings for the full data set. Each model has a native sample rate and window size, chosen independently of any of the datasets under consideration. Each audio sample is resampled to the model's native sample rate (though we experiment with alternatives in \ref{sec:audiomae_steelman}). When an example is shorter than the model's window size, we apply centered zero-padding to obtain the target length. When a model's window size is shorter than a target example, we frame the audio according to the model's window size, create an embedding for each frame, and then average the results. In the end, each example is associated with a single embedding vector. We then randomly choose a ﬁxed number $k$ of examples from each class, using a seeded random shuffle to ensure that the same training examples are used for every model. The $k$ examples are used to train a linear classifier over the pre-computed embeddings, and all remaining examples are used for evaluating the trained classifier. We use a binary cross entropy (BCE) loss, with sigmoid activation, and train the classifier to convergence. This process is repeated five times with different random seeds for each combination of model, dataset, and $k$, using the same set of five random seeds for each combination. We do this to report a reliable estimate of the classification performance~\citep{ghani2021randomized}. By default, we choose $k=32$.

For evaluation we compute (1) macro-averaged ROC-AUC (computing ROC-AUC for each class, and then averaging over all classes) and (2) Top-1 Accuracy. Reported metrics are averaged over the five training runs. 

\subsection{Experiment: Few-Shot Learning}
\label{sec:methods}

By varying the amount of training data, we can further understand the resilience of each embedding to reduced data availability. As it turns out, linear probes of pretrained embeddings is a surprisingly successful strategy for few-shot learning, as described in~\citep{chen2019closerfewshot}, which found this baseline method to be competitive with far more complicated approaches such as meta-learning. Other subsequent works have confirmed that `frustratingly simple' approaches with strong embeddings are sufficient for many few-shot learning tasks~\citep{frustrating2020}~\citep{tian2020}~\citep{hosoda2022done}. We leave comparison of more involved few-shot techniques for the future work.

We vary $k$ as powers-of-two between 4 and either 32 or 256, depending on the size of the target dataset. In the results, we provide results for all tested values of $k$ in Figure~\ref{fig:test_plots} and specifically for $k=32$ in Table~\ref{table:results}.

\subsection{Experiment: Visualizing Embedding Spaces}

We can also observe the geometry of the embedding space using a t-SNE transformation of the model embeddings~\citep{hinton2002stochastic}. The t-SNE transformation attempts to preserve distances in the embedding space while projecting to two dimensions.  In Fig.~\ref{fig:tsne} we plot t-SNE transforms for YAMNet, AudioMAE, BirdNET and Perch. Note that t-SNE plots can be tricky to interpret appropriately~\citep{wattenberg2016use}, though points which are close in the original space tend to be close after applying the t-SNE transform. We choose t-SNE over other alternatives specifically because it uses a shallow model, and we wish to preserve as closely as possible the structure of the embedding spaces.

\subsection{Experiment: Deep-Dive into AudioMAE}

In recent years, transformer and self-supervised models have taken a dominant position in machine learning research. Therefore, it may be surprising that AudioMAE - a self-supervised transformer - under-performed the humble EfficientNet-B1 architecture. We therefore performed a number of additional experiments to discover whether additional tweaking of the experimental setup would uncover hidden performance gains for the AudioMAE embeddings. We applied three different treatments on all six datasets, attempting to find a higher-performing operating point for the AudioMAE.

First, we compared quality of the pre-trained unsupervised embedding and the embedding obtained from supervised fine-tuning on AudioSet. Because the unsupervised objective is spectrogram reconstruction, one would expect that all relevant information should be present in the pre-trained embedding, but possibly suppressed by fine-tuning on the irrelevant AudioSet label-space. In fact, using the pre-trained or fine-tuned embedding does change the metrics, but not in a predictable way.

Second, we experimented with ignoring the audio sample rate when loading the target audio. Because the AudioMAE consumes 16kHz audio, any significant features above the Nyquist frequency of 8kHz will be lost when audio is resampled to the model's input rate. Instead of resampling, we may instead load the audio at its native sample rate and feed it directly to the model as though it were 16kHz.

Finally, we tried using a two-layer network with the pre-trained model, under the hypothesis that the raw self-supervised embedding may not be well aligned for classification tasks. The two-layer network consists of batch-normalization, a hidden layer with 2048 units (double the embedding dimensionality), a ReLU activation, and an output layer.

\subsection{Experiment: Varying Embedding Size}

We can view feature embeddings as a lossy compression of the input data. For instance, in terms of raw data, the embedding produced by the Perch model contains only 1.6\% of the data of the raw audio (a 1280-dimensional 32-bit float vector, derived from 5 seconds of 32 kHz audio encoded as 16-bit integers). Yet these embeddings enable efficient recognition of a wide range of global bird species. For this to work well, the classifier must learn features relevant to the classification problem while allowing irrelevant data to be discarded. This compression viewpoint suggests that embedding size may play a role in overall embedding quality.

We ran an additional ablation on embedding size while investigating the difference between BirdNET 2.2 and Perch models, which had embedding sizes 320 and 1280, respectively. We investigate the role of embedding size by comparing two versions of BirdNET with different embedding sizes (320 and 1024) and six versions of Perch (160, 320, 640, 960, 1280, 2560), using the same methodology and datasets described above.

Since we do not have ready access to the training infrastucture for the general audio models, we exclude them from this ablation.

\subsection{Experiment: No-Pretraining Baseline}

It is reasonable to ask how a custom-trained model would perform on the novel tasks we consider in this study. However, it is difficult to train a model from scratch with reasonable performance on the amount of data provided for the tasks. To provide this baseline, we extract MFCC features from the target audio and train a small two-layer neural network with 2048 hidden features on each dataset, mirroring a baseline considered in the BEANS benchmark~\citep{hagiwara2022beans}.

\subsection{Experiment: Comparison to a Regional Model}

One might also ask how transfer of the global bird embeddings compares to transfer of embeddings from a model trained on a smaller collection of bioacoustic classes. For this, we compare the transfer performance of a regional model trained for identification of 89 species in the Sierra Nevadas. 

\section{Models and Datasets}

\subsection{Model descriptions}
\label{sec:models}
We compare three models trained on bird data (BirdNET, Perch, and Sierras) to four models trained on variants of AudioSet~\citep{audioset} for general audio event detection (AudioMAE, PSLA, YAMNet, VGGish).

BirdNET and Perch are similar models, differing mostly in their training data. While Perch is trained exclusively on bird sounds data, BirdNET's training dataset also comprises of a relatively small fraction of non-birds sound data. %
\begin{table}[h!]
    \begin{center}
    \footnotesize
    \centering
    \begin{tabular}{lccccc}
    \toprule
                    & Architecture      &  Training Data & Window (s) & Embedding Size & CPU(ms/s)  \\
    \midrule
    Google Perch    & EfficientNet B1   & XenoCanto & 5.0 & 1280     & 24.3 \\
    BirdNET 2.2 / 2.3        & EfficientNet B1   & XC+ML+Custom & 3.0 & 320 / 1024 & 10.0 / 11.1 \\
    Sierras Birds   & EfficientNet B0   & XenoCanto+Custom & 5.0 & 1280     & 12.2 \\
    AudioMAE        & MAE (Large)     & AudioSet  & 10.0 & 1024 & 78.2 \\
    PSLA            & EfficientNet B2 & AudioSet  & 10.0 & 1408 & 246.0* \\
    YAMNet          & MobileNet v1      & AudioSet  & 0.96 & 1024 & 7.7 \\
    VGGish          & Modified VGG      & YouTube 8M   & 0.96 & 128 & 2.8 \\
    BEANS Baseline  & MFCCs + MLP       & N/A       & N/A & 160 & N/A \\
\bottomrule
    \end{tabular}
    \end{center}
    \caption{Summary of Embedding Model Characteristics. CPU(ms/s) is the benchmarked run-time for evaluating one audio window with the model, divided by model's window size. Models were benchmarked on a 4.3GHz AMD CPU with 12 cores. We expect that PSLA inference can be optimized to under 50ms.}
    \label{table:models}
\end{table}
AudioSet comprises an extensive compilation of over 2 million audio clips, each 10 seconds in duration. These clips are derived from YouTube videos and are categorically labeled according to the type of sound they contain, with a total of 527 unique classes. The classes include ‘wild animals,’ but the associated labels are very coarse (bird, frog, roaring cat) and constitute only about 2\% of the total dataset. To elaborate further, the specifications of the models are detailed as follows:   

 \textbf{Perch}\footnote{\url{https://tfhub.dev/google/bird-vocalization-classifier/4}} is an EfficientNet B1~\citep{tan2019efficientnet} trained on the full corpus of bird song recordings from Xeno-Canto (XC) downloaded in July, 2022. Because XC is weakly labeled (a single label for an entire file), we use an activity detector to select training windows from each file, as described in~\citep{denton2022improving}. During training we augment with MixUp~\citep{zhang2017mixup}, random gain adjustment, and random time-shifting of up to one second. The model is trained to classify all levels of the taxonomy for each recording simultaneously (species, genus, family, order). The base Google Perch model and further evaluation statistics are available at TFHub and supporting code is available on GitHub\footnote{\url{https://github.com/google-research/perch}}.   
 
 \textbf{BirdNET}~\citep{kahl2021birdnet} also uses an EfficientNet architecture, but does not use taxonomic outputs. BirdNET has a broader training set, including XC, the Macaulay Library, and labeled soundscape data from around the world, ultimately targeting many thousands of bird species. Additionally, BirdNET is trained to identify human speech, dogs, and many species of frogs. To enable a range of downstream use-cases, BirdNET trades off some accuracy for efficient computation. We report on BirdNET 2.2 and 2.3.  Version 2.3 features a higher embedding dimension (see Section~\ref{sec:embed_size}) and is trained on a larger bird species list. The BirdNET code is available on GitHub\footnote{\url{https://github.com/kahst/BirdNET-Analyzer}}, and includes support for training small classifiers on embeddings. 

\textbf{Sierra Birds}~\citep{denton2022improving} is a `regional' bird classification model, trained on 89 bird species found in the California Sierra Nevada mountains. The model uses an EfficientNet B0 architecture, marginally smaller than the B1 architecture used by the BirdNET and Perch models, and was primarily trained on XC recordings. The model training data was augmented with additional non-bird noise data, as described in~\citep{denton2022improving}.

\textbf{AudioMAE}~\citep{huang2022amae} is a more recent general audio model built with a transformer architecture. The model is trained on AudioSet with a self-supervision task, reconstructing masked spectrograms. The model consists of an encoder (which produces embeddings of patches of the spectrogram) and a decoder (which reconstructs the spectrogram from the patch embeddings). For this study, we use the embeddings produced by the encoder and discard the decoder. A 1024-dimensional embedding is obtained by averaging the per-patch embeddings, as is typical when using AudioMAE for classification tasks.  We evaluated a re-implementation of AudioMAE, using the `Large' model with 300M parameters, provided by Eduardo Fonseca~\citep{georgescu2022audiovisual}. This model obtains a mAP of 46.4 on AudioSet-2M after fine-tuning, comparable to the original AudioMAE's reported mAP of 47.3. We experimented with many configurations of AudioMAE, as described in Section~\ref{sec:audiomae_steelman}. AudioMAE training consists of a pre-training stage, where it is trained only for reconstruction of masked spectrograms, and a fine-tuning stage, where it is trained for supervised classification. None of these methods was consistently better than all others, so for brevity, we report results for the fine-tuned model with averaged embeddings unless otherwise noted. The original AudioMAE code can be accessed on GitHub\footnote{\url{https://github.com/facebookresearch/AudioMAE}}.

\textbf{PSLA}~\citep{gong2021psla} is an EfficientNet model trained on AudioSet, which we include for comparison with the EfficientNet-based global bird models. The model is competitive with SOTA models, obtaining a mAP score of 47.4 on Audioset-2M. Unlike the bird models, it uses an attention layer over the final embeddings to produce its predictions. We replace this attention layer with an averaging of the final embeddings to produce a summarized embedding for an audio segment, suitable for transfer learning.

\textbf{YAMNet} and \textbf{VGGish} are both convolutional models trained to predict AudioSet classes. YAMNet uses a MobileNetV1 architecture~\citep{howard2017mobilenets}. VGGish is an older audio event-detection model, using a variant of the VGG architecture and trained on an earlier version of AudioSet~\citep{vggish}. Both of these models process audio frames of 0.96 seconds. While the YAMNet model generates a feature embedding vector of 1024 dimensions, the VGGish embedding size is limited to 128 dimensions. The YAMNet\footnote{\url{https://github.com/tensorflow/models/tree/master/research/audioset/yamnet}} and VGGish\footnote{\url{https://github.com/tensorflow/models/tree/master/research/audioset/vggish}} codes can be accessed on GitHub. 

\textbf{BEANS Baseline} is not a pretrained model. Following the method in the BEANS Benchmark~\citep{hagiwara2022beans}, we extract a 160-channel MFCC representation of the input audio. This frequency information is then mean-pooled for the audio example, and a two-layer fully connected neural network is trained to predict the target class. This provides a simple baseline without pre-training for each task.

\subsection{Evaluation datasets}
We use a range of datasets for our analysis. These datasets were constructed by different groups with different goals and methodologies, and therefore vary in their characteristics. For instance, the RFCX and Watkins datasets contain cross-class contamination --- examples of a specific class where another unlabeled class is present. The bat species and Watkins datasets have variable clip length, whereas the other datasets have a fixed clip length. Table~\ref{table:datasets} presents an overview of all the datasets used in this work. 

\begin{table}[h!]
    \begin{center}
    \footnotesize
    \centering
    \begin{tabular}{lcccccc}
    \toprule
                    & \# Classes & Mean Class Size & Smallest Class & Sample Rate & Clip Length (s)  \\
    \midrule
    Godwit Calls                & 5 & 1343 & 628 & 44.1kHz & 3.0 \\
    Yellowhammer Dialects            & 2 & 772 & 444 & 48kHz & 3.5 \\
    Bats                     & 4 & 887 & 360 & 44.1kHz* & 1.0-13.0 \\
    Watkins Marine Mammals          & 32 & 60 & 35 & 22.05kHz & 0.1-10.0 \\
    RFCX Frog Species               & 12 & 50 & 37 & 48kHz & 5.0 \\
    RFCX Bird Species               & 13 & 53 & 34 & 48kHz & 5.0 \\
    \bottomrule
    \end{tabular}
    \end{center}
    \caption{Summary of target dataset characteristics. For the Bats dataset, frequency shifting was applied to move signal into the audible range. \\}
    \label{table:datasets}
\end{table}
 
 \textbf{Godwit Calls (GC):} The GC dataset contains five different calls of Black-tailed Godwit. The recordings were made by Ondrej Belfin as part of his masters thesis at the University of Groningen in the Netherlands~\citep{unknown2022thesis}. All recordings are 3 seconds long and are annotated by Ondrej Belfin himself. 

\textbf{Yellowhammer Dialects (YD):} The YD dataset comprises two dialects of Yellowhammer songs, denoted as X and B~\citep{petruskova2015review}, derived from audio recordings of Yellowhammer vocalizations. The two dialects are characterized based on variations of elements in the terminal phrase of the song. These recordings were sourced from submissions made through the BirdNET App, captured with various mobile phone microphones. Recordings were annotated in a two-step process. Connor Wood performed preliminary annotations, which were later refined by Pavel Pipek, a specialist in yellowhammer dialects at the Department of Ecology, Charles University in Prague. All recordings were acquired in 2020. Each audio recording within the data set has a duration of three seconds, facilitating a comprehensive analysis of the yellowhammer vocalizations. These dialects have a duration in the range 2.2-2.7 seconds and a fundamental frequency in the range of 5-6 kHz. 

 \textbf{Bats (BT):} The BT dataset contains four species of North American bats. The eastern red bat (\emph{Lasiurus borealis}, LABO) with 1,124 recordings, the little brown bat (\emph{Myotis lucifugus}, MYLU) with 1,119 recordings, the northern long-eared bat (\emph{Myotis septentrionalis}, MYSE) with 360 recordings, and the tricolored bat (\emph{Perimyotis subflavus}, PESU) with 948 recordings. The dataset is sourced from two origins: 1) Training dataset for NABat Machine Learning V1.0~\citep{gotthold2022training}, and 2) Dr. Patrick Wolff, US Army ERDC-CERL. The datasets were collected at ultrasonic sampling rates. We applied pitch shifting via sample rate conversion to these datasets, to bring the bat vocalizations into the audible range. After this pre-processing step, all audio has a sampling rate of 44.1 kHz. 
 
 \textbf{Watkins Marine Mammal Sounds Database (WMMSD):} The WMMSD dataset covers 60 species of marine mammals but we employ the `best of' category enlisted in the database as the species with higher quality and lower noise recordings. The taxonomical representation encompasses species from the Odontocete and Mysticete suborders within the order Cetacea, in addition to the Phocid and Otariid families, which are part of the clade Pinnipedia. The auditory documentation, spanning a substantial time period of seven decades, encapsulates a diverse range of recording methodologies, ambient acoustical conditions, and sampling frequencies~\citep{murphy2022residual}. The compilation of this auditory data was accomplished and annotated by several researchers including William Watkins, William Schevill, G. C. Ray, D. Wartzok, D. and M. Caldwell, K. Norris, and T. Poulte, and is openly accessible for academic use~\citep{sayigh2016watkins, watkins2021watkins}. The audio examples are cropped to the length of the actual vocalization, which means that the lengths of the audio files vary greatly by species. We exclude five classes for which there are a fewer than 32 examples provided, and two additional species which are characterized by very low frequency vocalizations (fin whale and northern right whale).

 \textbf{Rainforest Connection Kaggle dataset (RFCX Frogs, RFCX Birds):} This is the training data from the 2021 Species Audio Detection challenge, consisting of recordings of Puerto Rican birds and frogs. Both birds and frogs are present in the class list; to understand model performance on these taxa, we present results on each taxa separately, and all together.
    
The bird species in the RFCX data appear in the training data for both the Perch and BirdNET models, but most of these species have very limited training data. As of this writing, the median number of Xeno-Canto recordings for these thirteen species is just 17, and only two species have more than 50 recordings (the Bananaquit with 579 recordings, and the Black-Whiskered Vireo with 68 recordings). Thus, these are largely low-data species for these models.

\subsubsection{Dataset Limitations}
Each of the datasets we work with presents distinct difficulties. 

First, our methodology does not create an ideal train/test split when multiple examples originate from the same original recording. Ideally, different source recordings or entire recording sites would appear consistently as train or test data to reflect model generalization to new conditions. We do not have sufficient metadata available for all datasets to perform such a split, and so results may overestimate model generalization on the target tasks. Instead, we treat each example independently, and create a train/test split over the examples we have. We believe this issue affects only the Bats, RFCX, and a subset of the Watkins species.

Secondly, some recordings contain additional unlabeled vocalizations, which may lead to under-estimation of model quality. This is especially the case for the Watkins and RFCX frog datasets. (See Table~\ref{table:marine_confusion} for some analysis of the Watkins dataset.)

\section{Results}

\subsection{Linear Probes on Feature Embeddings}

\begin{figure*}[b!]
    \centering
    \includegraphics[width=\textwidth]{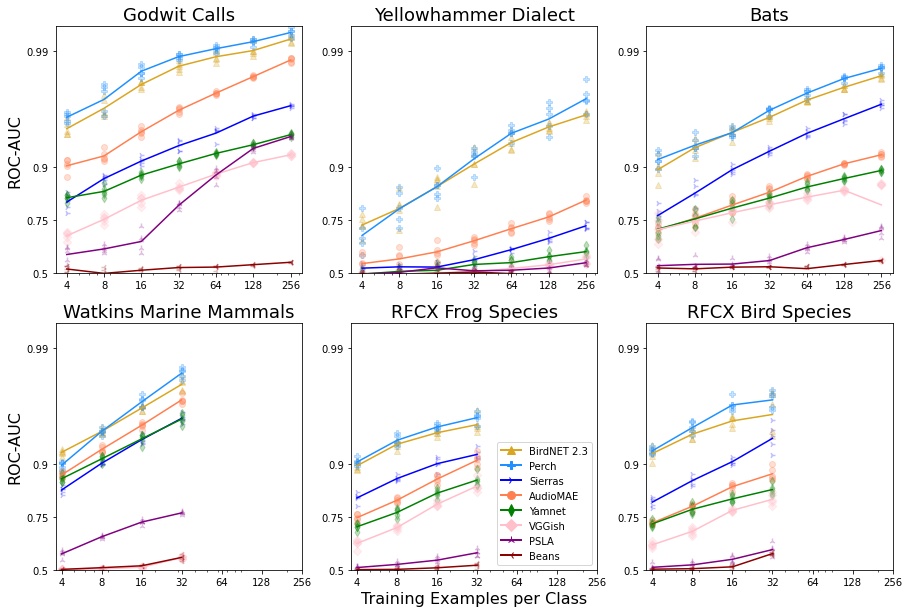}
    \caption{Results of Few-Shot Transfer Learning Tasks. ROC-AUC scores are plotted with log-odds scaling. A point is plotted for each experiment, and the curve connects the average quality for experiments at each number of training examples per class.}
    \label{fig:test_plots}
\end{figure*}

Our study delves into the classification performance by employing a variety of embeddings with linear probes for novel bioacoustic tasks. For an in-depth look, refer to Table~\ref{table:results}, which elucidates the results when training linear probes with $k=32$ examples per class. Additionally, Figure~\ref{fig:test_plots} provides a visual representation of the results across diverse training data sizes, ranging from 4 to either 32 or 256 examples per class, contingent on the dataset size.

\begin{table}[h!]
    \begin{center}
    \footnotesize
    \centering
    \setlength\tabcolsep{5pt} 
    \begin{tabular}{lccccccccccccccc}
    \toprule 
Model		 & \multicolumn{2}{c}{GC} & \multicolumn{2}{c}{YD} & \multicolumn{2}{c}{BT} & \multicolumn{2}{c}{WMMSD} & \multicolumn{2}{c}{RFCX Frogs} & \multicolumn{2}{c}{RFCX Birds} \\
			 & Top-1 & AUC & Top-1 & AUC & Top-1 & AUC & Top-1 & AUC & Top-1 & AUC & Top-1 & AUC \\
\hline
Perch            & \bf 0.92 & \bf 0.99 & \bf 0.87 & \it 0.91 &  \bf 0.86 & \bf 0.97 & \bf 0.83 & \bf 0.98 & \bf 0.74 & \it 0.96 & \bf 0.83 & \bf 0.97 \\
BirdNET 2.3      &  0.91 &  0.99 &  0.84 &  0.91 &  0.85 &  0.96 &  0.81 &  0.98 &  0.73 &  0.95 &  0.78 &  0.96 \\
Sierras          &  0.76 &  0.93 &  0.56 &  0.57 &  0.77 &  0.93 &  0.72 &  0.96 &  0.65 &  0.92 &  0.69 &  0.94 \\
\hline
AudioMAE         &  0.85 &  0.96 &  0.61 &  0.66 &  0.63 &  0.85 &  0.74 &  0.96 &  0.56 &  0.89 &  0.43 &  0.85 \\
PSLA             &  0.20 &  0.80 &  0.65 &  0.51 &  0.27 &  0.57 &  0.06 &  0.76 &  0.10 &  0.59 &  0.06 &  0.60 \\
YamNet           &  0.71 &  0.91 &  0.54 &  0.55 &  0.61 &  0.83 &  0.69 &  0.96 &  0.48 &  0.86 &  0.43 &  0.84 \\
VGGish           &  0.63 &  0.86 &  0.51 &  0.51 &  0.57 &  0.80 &  0.04 &  0.56 &  0.48 &  0.85 &  0.39 &  0.81 \\
\hline 
BEANS Baseline           &  0.14 &  0.53 &  0.72 &  0.51 &  0.24 &  0.53 &  0.04 &  0.56 &  0.11 &  0.52 &  0.05 &  0.58 \\
    \bottomrule
    \end{tabular}
    \end{center}
    \caption{Table of Results. We report the top-1 accuracy and ROC-AUC score of the linear classifiers, averaged over five runs, for each data set. All results are for 32 training examples per species. Entries are bold-faced if the model scored highest on all five runs, and italic if highest on four of five runs. \\}
    \label{table:results}
\end{table}

The Perch and BirdNET 2.3 models obtain similar performance. However, Perch achieved the highest Top-1 accuracy and AUC across all the datasets, making it the most consistent performer. It performed particularly well with ``Godwit Calls'' and ``Bat Species'', with AUCs of 0.99 and 0.97, respectively. Similarly, BirdNET 2.3 exhibited a good performance, especially with Godwit Calls (GC) and Bat species (BT) (0.99, 0.96). 

Both bird models significantly outperform the AudioSet models on all tasks (VGGish, YAMNet, PSLA, and AudioMAE). The macro-averaged ROC-AUC scores are typically high, suggesting good binary classification on each class individually. In case of AudioMAE, the performance dropped significantly, especially noticeable with the Yellowhammer dialects (YD) and RFCX birds datasets, which had lower AUCs of 0.66 and 0.78, respectively. The performance declined further using the YamNet model. The Top-1 accuracy was relatively low across datasets, and AUCs were significantly lower, particularly for the YD dataset. The VGGish model had the lowest performance across all datasets, notably underperforming on the WMMSD dataset with a very low Top-1 accuracy of 0.04 and AUC of 0.52. In summary, the Perch and BirdNET 2.3 models outperformed the others in terms of both Top-1 accuracy and AUC, demonstrating superior generalizability across various bioacoustic datasets. On the other hand, VGGish showed the weakest performance. Among the three models trained on the AudioSet, the transformer-based AudioMAE model outperformed the CNN-based VGGish, YAMNet, and PSLA models across all datasets except for the RFCX Birds dataset in which YAMNet performed slightly better. The performance gain for YD and GC datasets was significant.

\subsection{Experiment: Few-shot learning}
In Fig.~\ref{fig:test_plots} we show results with varying amounts of training data per class. We again find that using transfer learning with global bird models (BirdNET and Perch) consistently outperforms general event-detection models trained on YouTube data (AudioMAE, Yamnet, and VGGish).

In all cases, the bird models have an ROC-AUC significantly greater than 0.5 even with only 4 training examples.

\subsection{Visualizing Embedding Spaces}

\begin{figure*}[h!]
    \centering
    \begin{subfigure}[b]{0.24\textwidth}
        \includegraphics[width=\textwidth]{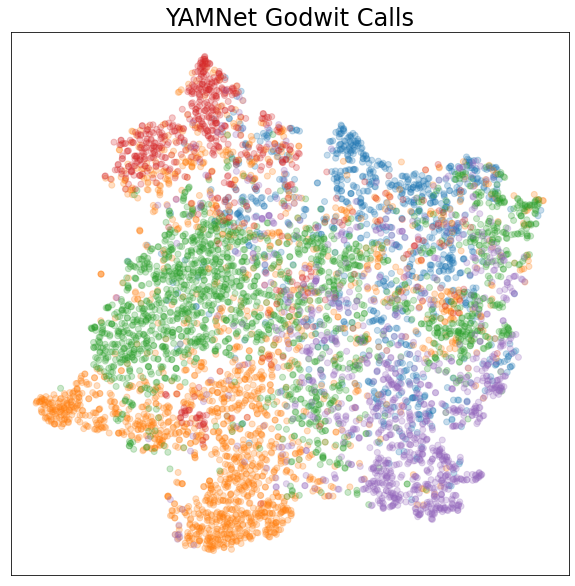}
    \end{subfigure}
    \begin{subfigure}[b]{0.24\textwidth}
        \includegraphics[width=\textwidth]{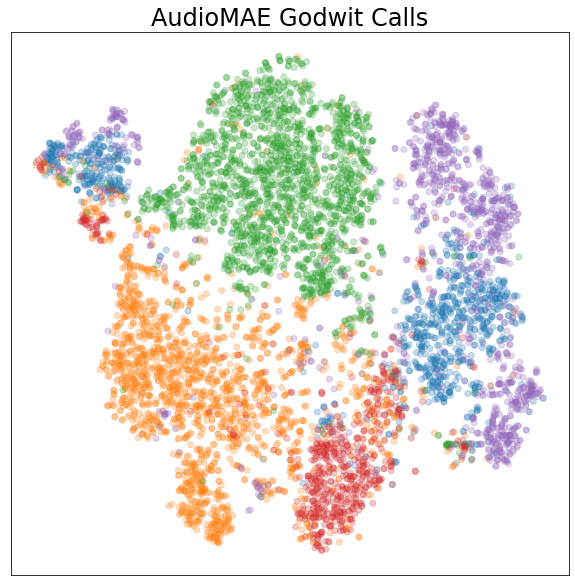}
    \end{subfigure}
    \begin{subfigure}[b]{0.24\textwidth}
        \includegraphics[width=\textwidth]{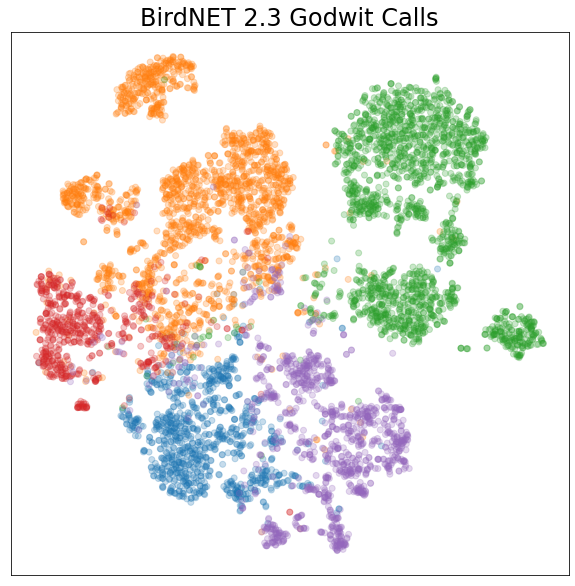}
    \end{subfigure}
    \begin{subfigure}[b]{0.24\textwidth}
        \includegraphics[width=\textwidth]{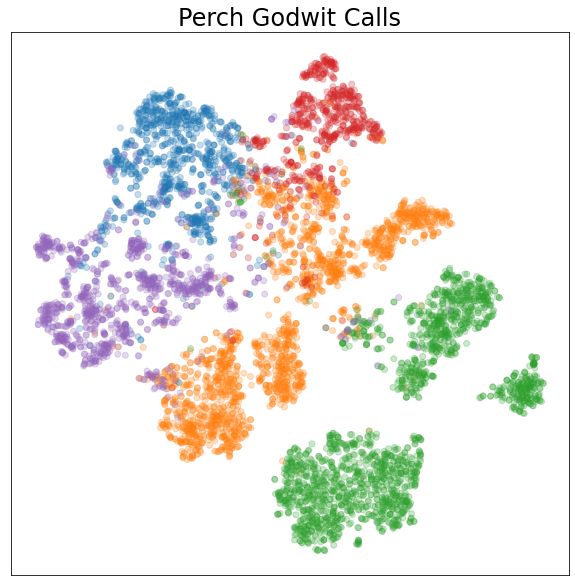}
    \end{subfigure}
    \begin{subfigure}[b]{0.24\textwidth}
        \includegraphics[width=\textwidth]{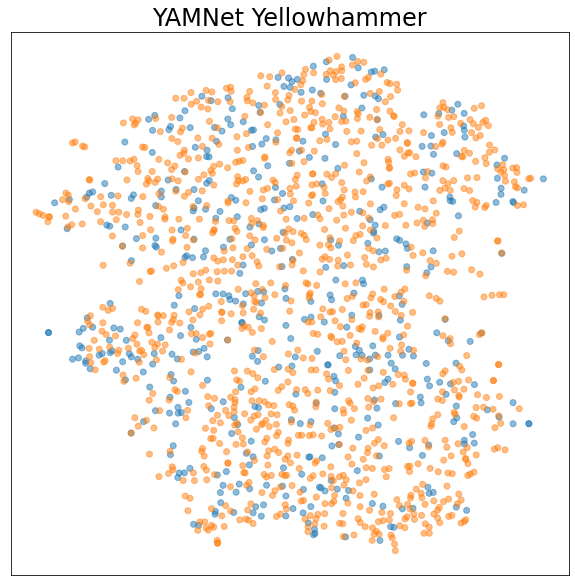}
    \end{subfigure}
    \begin{subfigure}[b]{0.24\textwidth}
        \includegraphics[width=\textwidth]{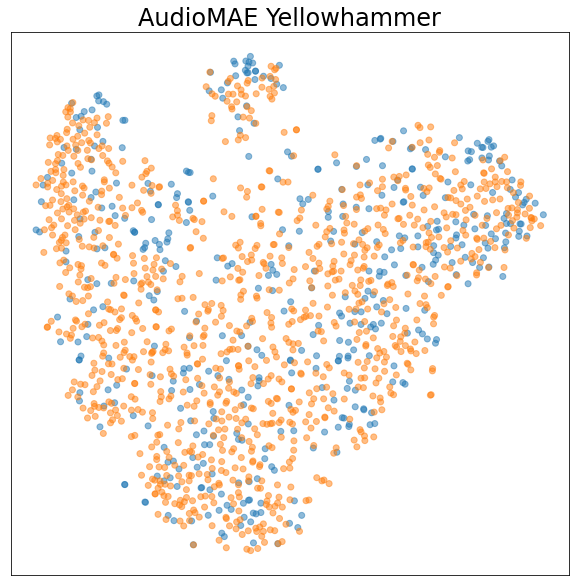}
    \end{subfigure}
    \begin{subfigure}[b]{0.24\textwidth}
        \includegraphics[width=\textwidth]{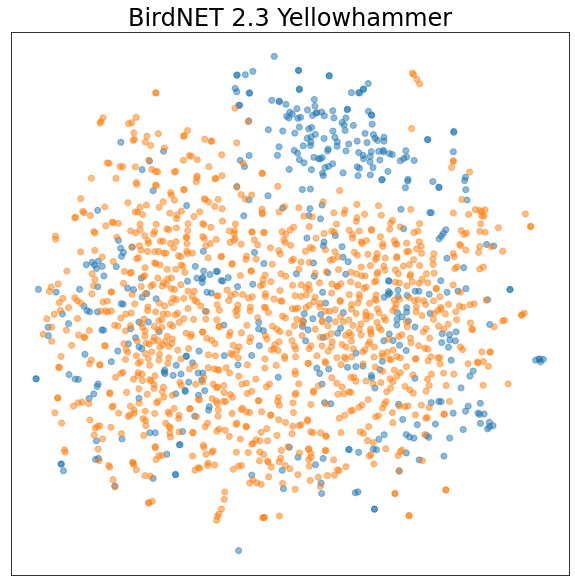}
    \end{subfigure}
    \begin{subfigure}[b]{0.24\textwidth}
        \includegraphics[width=\textwidth]{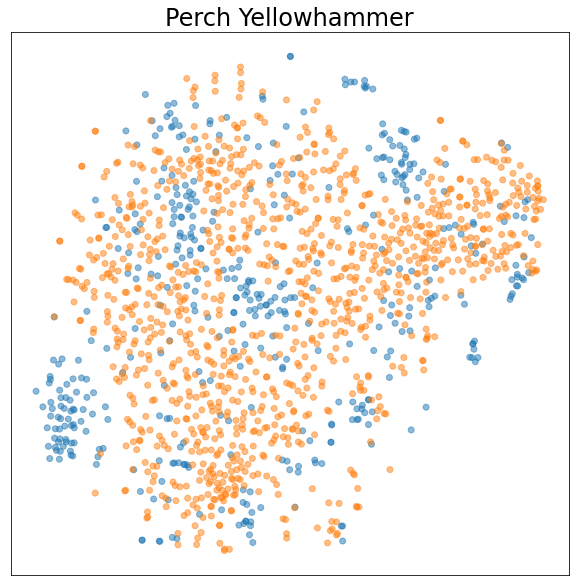}
    \end{subfigure}
    \caption{t-SNE plots of Godwit and Yellowhammer embeddings. Points are colored by class.}
    \label{fig:tsne}
\end{figure*}

In the easier Godwit problem (Fig.~\ref{fig:tsne}), we observe cleaner clustering of labeled data in the Perch embeddings, with large margins suggesting easy linear separability of the classes. By contrast, there are no clean margins between classes in the YAMNet embeddings, and smaller, noisier margins for the AudioMAE embeddings. 

For the more difficult Yellowhammer problem, we observe a complete intermixing of the two classes for YAMNet, explaining the model's inability to linearly separate the classes. For AudioMAE, which performs marginally better, we can observe a couple pockets of concentrated blue points, but no clear clustering. For Perch and BirdNET we see some clustering, but still a great deal of inter-mixed data, explaining the slow improvement on this task.

\subsection{Experiment: Deep-Dive into AudioMAE}
\label{sec:audiomae_steelman}

The best overall AudioMAE performance was obtained by using a 2-layer perceptron and no audio resampling with the pre-trained embeddings. Despite substantial effort, we found the global bird embeddings - with no additional tweaking - uniformly outperformed the AudioMAE model. PSLA benefitted significantly from a two-layer probe, but still underperformed both AudioMAE and the global bird classifiers.

\begin{table}[h!]
    \begin{center}
    \footnotesize
    \centering
    \setlength\tabcolsep{4pt} 
    \begin{tabular}{lccccccccccccccccc}
    \toprule 
Model		 & Probe & RS? & \multicolumn{2}{c}{GC} & \multicolumn{2}{c}{YD} & \multicolumn{2}{c}{BT} & \multicolumn{2}{c}{WMMSD} & \multicolumn{2}{c}{RFCX-F} & \multicolumn{2}{c}{RFCX-B} \\
			 & & & Top-1 & AUC & Top-1 & AUC & Top-1 & AUC & Top-1 & AUC & Top-1 & AUC & Top-1 & AUC \\
\hline
Perch            & LR & Y &  0.92 &  0.99 & \bf 0.87 &  0.91 &  0.86 &  0.97 & \bf 0.83 &  \bf 0.98 &  0.74 &  0.96 &  0.83 &  0.97 \\
Perch            & LR & N &  0.91 &  0.99 &  0.86 &  \bf 0.93 &  0.80 &  0.94 &  0.80 &  0.98 & \bf 0.76 &  0.96 &  0.83 &  0.98 \\
Perch            & 2LP & Y & \bf 0.92 &  0.99 &  0.87 &  0.92 & \bf  0.86 & \bf  0.97 &  0.81 &  0.98 &  0.73 &  0.96 &  \bf 0.83 &  0.97 \\
Perch            & 2LP & N &  0.92 & \bf  0.99 &  0.84 &  0.92 &  0.81 &  0.95 &  0.79 &  0.96 &  0.76 &  \bf 0.96 &  0.83 &  \bf 0.98 \\
\hline
BN2.3      & LR & Y &  0.91 &  0.99 &  0.84 &  0.91 &  0.84 &  0.96 &  0.81 &  0.98 &  0.73 &  0.95 &  0.78 &  0.96 \\
\hline
MAE/p          & LR & Y &  0.72 &  0.91 &  0.57 &  0.60 &  0.68 &  0.88 &  0.60 &  0.93 &  0.59 &  0.91 &  0.53 &  0.90 \\
MAE/p          & LR & N &  0.80 &  0.95 &  0.59 &  0.62 &  0.65 &  0.87 &  0.68 &  0.95 &  0.64 &  0.93 &  0.65 &  0.94 \\
MAE/p          & 2LP & Y &  0.82 &  0.96 & \it 0.66 & 0.64 & \it 0.74 & \it 0.91 &  0.78 &  0.97 &  0.69 &  0.94 &  0.64 &  0.94 \\
MAE/p          & 2LP & N &  0.84 &  0.97 &  0.63 &  0.65 &  0.73 &  0.91 &  \it 0.81 &  \it 0.97 &  \it 0.70 &  \it 0.94 &  \it 0.72 &  \it 0.96 \\
\hline
MAE/f          & LR & Y &  0.85 &  0.96 &  0.61 &  0.66 &  0.63 &  0.85 &  0.74 &  0.96 &  0.56 &  0.89 &  0.43 &  0.85 \\
MAE/f          & LR & N &  0.86 &  0.97 &  0.62 &  0.66 &  0.61 &  0.84 &  0.76 &  0.97 &  0.62 &  0.91 &  0.55 &  0.88 \\
MAE/f          & 2LP & Y &  0.85 &  0.97 &  0.63 &  0.67 &  0.65 &  0.86 &  0.79 &  0.97 &  0.60 &  0.89 &  0.49 &  0.86 \\
MAE/f          & 2LP & N &  \it 0.87 & \it 0.97 &  0.64 & \it 0.67 &  0.64 &  0.86 &  0.80 &  0.97 &  0.60 &  0.89 &  0.62 &  0.90 \\
    \hline
PSLA             &  LR   & N &  0.20 &  0.80 &  0.65 &  0.51 &  0.27 &  0.57 &  0.06 &  0.76 &  0.10 &  0.59 &  0.06 &  0.60 \\
PSLA             &  2LP  & N &  0.82 &  0.96 &  0.46 &  0.58 &  0.46 &  0.76 &  0.63 &  0.94 &  0.35 &  0.77 &  0.40 &  0.81 \\
    \bottomrule
    \end{tabular}
    \end{center}
    \caption{Results of AudioMAE Steel-man Experiments. Results are all for 32 examples per class. Probe is LR for linear regression or 2LP for two-layer perceptron. `RS' indicates whether the audio was resampled to the embedding model's preferred sample rate. `MAE/p' is the pretrained unsupervised embedding model, and `MAE/f' is fine-tuned with supervision on AudioSet. The highest score in each column is bold-faced, and the highest AudioMAE score is in italic.}
    \label{table:mae_ablation}
\end{table}

\subsection{Experiment: Varying Embedding Size}
\label{sec:embed_size}

\begin{figure*}[b!]
    \centering
    \includegraphics[width=\textwidth]{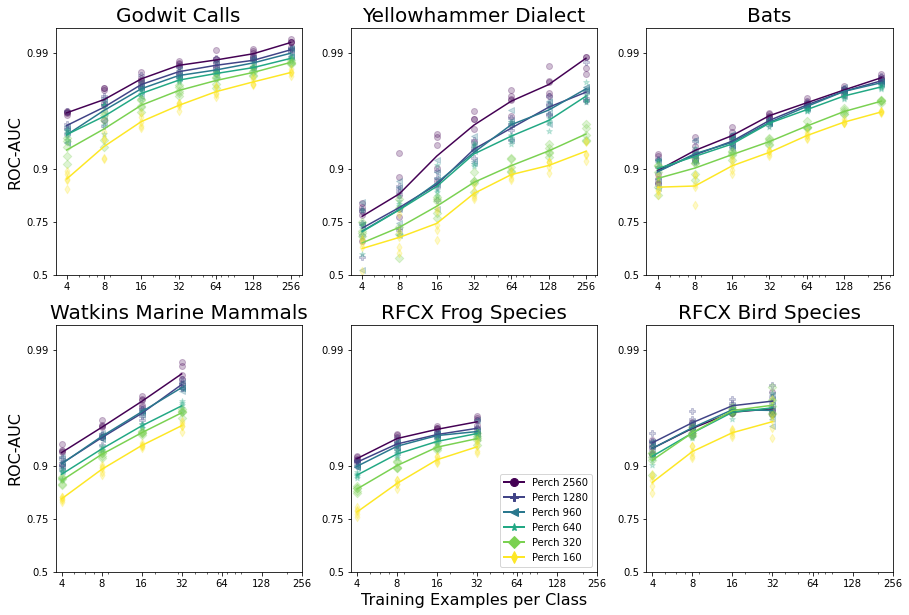}
    \caption{Results of Embedding Size Ablation Test. ROC-AUC scores are plotted with log-odds scaling.}
    \label{fig:embedding_size_plots}
\end{figure*}

An ablation over the embedding dimension is summarized in Fig.~\ref{fig:embedding_size_plots}. The Top-1 Accuracy and ROC-AUC scores on different datasets using various embedding sizes are shown in Table~\ref{table:size_ablation}. Perch with a 320-dimensional embedding (matching BirdNET 2.2) has significantly degraded quality in all tasks. Doubling the base Perch embedding dimension to 2560 yields a further increase in model performance for some downstream tasks. 

Increasing the size of the BirdNET embedding to 1024 led to similar performance as the Perch model in most downstream tasks.

\begin{table}[h!]
    \begin{center}
    \footnotesize
    \centering
    \setlength\tabcolsep{4pt} 
    \begin{tabular}{lccccccccccccccc}
    \toprule 
Model		 & Size & \multicolumn{2}{c}{Godwit Calls} & \multicolumn{2}{c}{Yellowhammer} & \multicolumn{2}{c}{Bat Species} & \multicolumn{2}{c}{Watkins} & \multicolumn{2}{c}{RFCX Frogs} & \multicolumn{2}{c}{RFCX Birds} \\
			 &   & Top-1 & AUC & Top-1 & AUC & Top-1 & AUC & Top-1 & AUC & Top-1 & AUC & Top-1 & AUC \\
\hline
Perch            & 2560 &  0.91 &  0.99 &  0.92 &  0.96 &  0.85 &  0.96 &  0.83 &  0.98 &  0.75 &  0.96 &  0.82 &  0.97 \\
Perch            & 1280 &  0.91 &  0.99 &  0.88 &  0.93 &  0.85 &  0.96 &  0.80 &  0.98 &  0.73 &  0.95 &  0.83 &  0.97 \\
Perch            & 960 &  0.91 &  0.98 &  0.88 &  0.93 &  0.85 &  0.96 &  0.80 &  0.98 &  0.74 &  0.95 &  0.82 &  0.97 \\
Perch            & 640 &  0.90 &  0.98 &  0.87 &  0.92 &  0.84 &  0.96 &  0.74 &  0.97 &  0.73 &  0.95 &  0.81 &  0.97 \\
Perch            & 320 &  0.89 &  0.98 &  0.80 &  0.87 &  0.80 &  0.94 &  0.71 &  0.96 &  0.71 &  0.94 &  0.81 &  0.97 \\
Perch            & 160 &  0.88 &  0.97 &  0.80 &  0.84 &  0.79 &  0.93 &  0.66 &  0.95 &  0.68 &  0.93 &  0.78 &  0.96 \\
\hline 
BirdNET 2.3      & 1024 &  0.91 &  0.99 &  0.84 &  0.91 &  0.85 &  0.96 &  0.81 &  0.98 &  0.73 &  0.95 &  0.78 &  0.96 \\
BirdNET 2.2      & 320 &  0.90 &  0.98 &  0.83 &  0.88 &  0.83 &  0.95 &  0.79 &  0.98 &  0.75 &  0.96 &  0.79 &  0.96 \\
    \bottomrule
    \end{tabular}
    \end{center}
    \caption{Results of Embedding Size Ablation. We report the top-1 accuracy and ROC-AUC score of the linear classifiers, averaged over five runs, for each data set. All results are for 32 training examples per species. All Perch models were trained for this ablation from scratch.}
    \label{table:size_ablation}
\end{table}

\subsection{Experiment: No-Pretraining Baseline and Regional Model}

The results for the BEANS MFCC baseline and Sierra Birds model are included in Table~\ref{table:results} and Figure~\ref{fig:test_plots}. We find that the MFCC model barely exceeds chance, significantly underperforming transfer from VGGish, the weakest of the audio domain models.

We find that the global bird models significantly outperform the `regional' Sierra Birds model. Furthermore, the regional model underperforms general audio embeddings on some datasets.

\section{Discussion}


The performance results displayed in Fig.~\ref{fig:test_plots} underscore the value of transfer learning with global bird models such as BirdNET and Perch. These models consistently outperformed general event-detection models trained on broader auditory data, such as YouTube-sourced data utilized by AudioMAE, PSLA, YAMNet, and VGGish. This observation is pivotal as it suggests that models specifically trained on bird data possess a heightened capacity for generalization, successfully identifying and analyzing previously not encountered bioacoustic patterns. 

In our study, we employed a linear probe method to evaluate the pre-trained embeddings of various models, including BirdNET and Perch. This method, essentially a linear layer placed on top of the selected embeddings, serves as a tool for assessing the availability of desired task-specific information within these embeddings. Consequently, our findings, as demonstrated in Fig.~\ref{fig:test_plots} and Table~\ref{table:results}, suggest that the embeddings derived from global bird models contain rich, task-specific information conducive to bioacoustic analysis. This is in contrast to general auditory event-detection models, which although trained on a broader spectrum of auditory data, do not exhibit the same level of performance. This superior generalization capability of deep embeddings from global bird models is an important finding, as it highlights the potential of these specialized models in providing a more robust and adaptable framework for varied bioacoustic tasks by learning good quality embeddings from data. In the realm of bioacoustic sound event detection, the ability to generalize across distinct taxonomic categories and acoustic characteristics is invaluable, as it facilitates the fine-grained classification of call types, song dialects, and out-of-scope identification of acoustic events.  

This finding might be attributed to the inherent diversity and complexity found in bird vocalizations. Bird songs and calls occupy a broad range both temporally and in the spectral domain, exhibiting diverse frequency modulations, harmonic structures, and rhythmic patterns. This wide array of acoustic characteristics provides a rich and versatile training data set for models such as BirdNET and Perch. The comprehensive nature of these vocalizations may have facilitated the models' ability to learn more generalized representations of bioacoustic patterns. 

This versatility in bird vocalizations has a dual implication. Firstly, it enriches the training dataset, providing varied instances for the model to learn from, and subsequently, it enables the model to capture a broader range of acoustic patterns, improving its ability to generalize to novel categories. Secondly, the acoustic diversity among bird species might mimic the bioacoustic variability encountered in other taxa, thus further enhancing the model's generalization capabilities when applied to sounds from different taxa. Indeed, previous work has found a collection of mechanisms, termed MEAD, by which all birds and mammals vocalize~\citep{elemans2015universal}, and which may help explain the successful transfer of bird features to other taxa. This hypothesis provides an intriguing direction for future research -- exploring the specific characteristics of bird vocalizations that contribute to these superior generalization capabilities. Understanding these characteristics could guide the collection and selection of training data for future bioacoustic models, with the aim of maximizing their generalization potential. 

The extensive diversity inherent in bird vocalizations, both in terms of acoustic characteristics and species diversity, is not just a theoretical advantage but also a practical one. The availability of a vast array of bird species audio data provides an advantageous basis for model training. Recent work has shown that training on a larger number of diverse classes helps model generalization more than adding data from a smaller number of classes~\citep{luo2023closer}. To make this concrete, we include a comparison to a regional bird classification model trained on 89 bird species from the Sierra Nevadas. The global bird models, which were trained on a larger variety of classes, exhibited superior performance compared to the Sierra Birds model. This underperformance of the regional model is further highlighted when compared to general audio embeddings, where it falls short in some datasets. This aligns with findings from the above paper which emphasized that a greater number of training classes enhances the generalization ability of models. The regional model's limited scope in terms of training classes could be a key factor in its lower effectiveness compared to the more broadly trained global bird models.  

We designed an experiment aimed to uncover any potential performance gains hidden within the architecture of the AudioMAE, which is SOTA and widely believed to have a superior architecture. Despite the general dominance of transformer and self-supervised models in machine learning in the recent years, AudioMAE was outperformed by the more basic EfficientNet-B1. The investigation involved three key experiments: examining the quality of pre-trained vs. fine-tuned embeddings, bypassing audio sample rate adjustments, and employing a two-layer network to align the self-supervised embedding for classification tasks. Surprisingly, the best performance of AudioMAE was achieved using the pre-trained embeddings with a two-layer perceptron and without audio resampling. However, even with these adjustments, the global bird embeddings, without any tweaking, consistently outperformed the AudioMAE model. We also found that PSLA, a near-SOTA AudioSet model with an EfficientNet architecture, performed poorly on our target tasks. Together, these findings demonstrate that the data effects dominate the architecture effects.

To demonstrate the robustness of each embedding in the context of reduced data availability, we ran experiments to train the linear probes by varying the amount of training data. Our results, as seen in Fig.~\ref{fig:test_plots}, have shown promising prospects for bioacoustic recognition tasks even when faced with as little as 4 training samples. The global bird models once again outperform other models in this experiment. This shows that these models can be used for active learning on novel tasks, starting even from a handful of examples. 
\begin{table}[!t]
    \begin{center}
    \footnotesize
    \centering
    \setlength\tabcolsep{5pt} 
    \begin{tabular}{llc}
    \toprule
Species & Confused Species & Confusion Rate \\
    \midrule
Bearded Seal                     & Bowhead Whale                    & 0.186 \\
Pantropical Spotted Dolphin      & Spinner Dolphin                  & 0.097 \\
Common Dolphin                   & Striped Dolphin                  & 0.091 \\
Frasers Dolphin                  & Pantropical Spotted Dolphin      & 0.082 \\
Killer Whale                     & Narwhal                          & 0.067 \\
    \bottomrule
    \end{tabular}
    \end{center}
    \caption{Top five marine mammal species confusions, averaged over five runs with the Perch model, using 32 examples per class. Bearded Seal and Bowhead Whale often appear in the same recording, though only one is labeled. \\}
    \label{table:marine_confusion}
\end{table}
Lower Top-1 accuracy scores (see Table~\ref{table:results}) suggest that inter-class calibration may still be a difficulty for simple linear probes, though \emph{unlabelled vocalizations} in the test set may account for some difficulty. For the Watkins dataset, a significant amount of confusions (18.6\%) occurred between bearded seals and bowhead whales, two highly vocal Arctic marine mammal species (see Table~\ref{table:marine_confusion}). Both species are known to overlap in range and are frequently recorded together, especially during the late spring and early summer months~\citep{chou2020seasonal}. This is also the case for the weakly-labeled training data we used, which explains the comparatively high degree of confusion. More sophisticated pre-processing of the training data and adding some strongly labeled data would help to increase the classification performance for these two species. The confusion between co-occurring dolphin species is also not surprising. First, these data were downsampled to the audible frequency range, which will cutoff higher frequency components of the vocalizations. In addition, dolphin species are generally difficult to classify acoustically~\citep{rankin2017acoustic} because they produce highly variable vocalizations including whistle, echolocation clicks, and burst pulses. Lastly, dolphins also occur in mixed species groups which can make it challenging to obtain clean training data.

We also see a particularly high variance in model quality for the YD dataset in the low-data regime. Since this is only a two-class problem, there are fewer total examples used for training in the low-data regime. However, this is also a subtle problem: The Yellowhammer dialect is distinguished by the order of the last two notes of the song: mid-then-high versus high-then-mid. Other variations in timbre of the initial portion of the song and up- or down-sweep in the high note do not distinguish between the two dialects. The subtlety of the problem apparently makes it easy to over-generalize from few examples. 

To demonstrate the efficacy of embeddings as condensed 'fingerprints' of raw audio, we conducted experiments that revealed an enhancement in separability for bioacoustic tasks with the increase in embedding size. However, this improvement comes at the cost of increased model size and reduced inference speed. Newer versions of BirdNET have a larger embedding size as a result of these observations.

\section{Conclusions}

Our study explored generalizablility of feature embeddings within the bioacoustics domain, focusing on the application of large-scale audio classification models to previously unencountered taxonomic groups such as marine mammals, bats, and frogs, in addition to intraspecific calls and dialects of a bird species. Our study supports the hypothesis that feature embeddings, especially those derived from bird data, can effectively represent high-dimensional categorical or discrete features as a low-dimensional continuous vector space.  We report that embeddings derived from models specifically trained using bird sounds data consistently facilitated superior classification quality compared to those trained on broader audio datasets.  This study's findings also suggest that feature embeddings from global bird acoustic classifiers can be effectively utilized for few-shot transfer learning. This enables the acquisition of new classes with a minimal amount of training data. Our empirical findings have significant implications for Passive Acoustic Monitoring, potentially enhancing the methods by which we detect and classify animal species based on their sounds. This could revolutionize the application of PAM, particularly in low-data regimes, by enabling more effective transfer learning between coarse-level classification and more fine-grained vocalization classification.

\section{Acknowledgements}


The German Federal Ministry of Education and Research is funding the development of BirdNET through the project ``BirdNET+'' (FKZ 01|S22072). Additionally, Federal Ministry of Environment, Nature Conservation and Nuclear Safety is funding the development of BirdNET through the project ``DeepBirdDetect'' (FKZ 67KI31040E). BirdNET is also supported by Jake Holshuh (Cornell class of ’69) and The Arthur Vining Davis Foundations. The work in the K. Lisa Yang Center for Conservation Bioacoustics is made possible by the generosity of K. Lisa Yang to advance innovative conservation technologies to inspire and inform the conservation of wildlife and habitats. BG acknowledges funding from the Landesforschungsförderung Hamburg within the AuTag BeoFisch (LFF-FV91) project while working as a researcher at the Hamburg University of Applied Sciences. We also acknowledge the rest of the Google Perch team (Vincent Dumoulin, Eleni Triantafillou, Bart van Merrienboer, and Jenny Hamer) who contributed to the development of the Perch model and offered numerous helpful suggestions on this work.
\bibliographystyle{unsrtnat}
\bibliography{references}

\end{document}